# Dark matter candidates: black holes
# and gravitationally bound black hole atoms in n-space

**Mario Rabinowitz**

Armor Research, 715 Lakemead Way, Redwood City, CA 94062-3922

E-mail: Mario715@earthlink.net

**Abstract**

A comparison is made of a range of proposals to include free and bound black holes as either a small component or the dominant constituent of dark matter. The hypothesis that dark matter consists of atomic gravitationally bound primordial black holes is closely examined in 3-space, as well as in n-space. It is demonstrated that for dimensions greater than 3, stable gravitational or electrostatic atoms cannot be bound by energy constraints. The total number of degrees of freedom of a d-dimensional body in n-space is derived so that equipartition of energy may be applied in the early universe. Blackbody and Hawking radiation are generalized to n-space. A promising new approach to black hole entropy is presented.

Mario Rabinowitz
Armor Research
715 Lakemead Way
Redwood City, CA 94062-3922
e-mail: Mario715@earthlink.net
Phone & FAX 650,368-4466
Manuscript: 16 pages
March  , 2003

## 1. Introduction

Black holes in differing scenarios have long been considered as dark matter candidates. More recently, Chavda and Chavda have proposed that dark matter consists of stable gravitationally bound primordial black holes with quantized orbits, which they call holeum[1]. For them the early universe does not allow Hawking radiation, and for later times they argue by analogy with a neutron that holeum will not Hawking decay (radiate), just as a bound neutron will not decay even though a free neutron does decay. Other proposals have also been made for free little black holes (LBH) and LBH in gravitationally bound atoms (GBA) [2, 3] as constituents of dark matter. Because of its potential to solve an important long-standing problem in astrophysics, GBA will be analyzed here in much closer detail than has been previously done.

Gravitational atoms have quantized orbits for the same reason that the orbits of ordinary electrostatic atoms are quantized. This follows directly from the quantization of angular momentum. Because they relate their analysis [1] to the early universe, it is appropriate to extend the analysis to macroscopic higher dimensions to which some theories attribute physical reality, and to see if n-space strengthens the case for GBA.

## 2. Quantized Gravitational Orbits in n-Space

Let us consider quantized non-relativistic gravitational circular orbits in n-space, where n = 3, 4, 5, ... □ is the number of spatial dimensions in the space-time manifold of (n+1) dimensions. They are an analog of electrostatic atomic orbitals. Ordinary matter does not have a density high enough to make such orbits achievable, but LBH do. For example, in 3-space, a $10^{-5}$ kg LBH with a radius of ~ $10^{-32}$ m has a density of ~ $10^{90}$ kg/$m^3$ , whereas nucleon densities are only ~ $10^{18}$ kg/$m^3$.

The results of previous derivations [2, 4] for quantized circular gravitationally bound orbits for the orbital radius, orbital velocity, binding energy, and the discrete gravitational spectrum (identical in form to the electromagnetic spectrum of the hydrogen atom) are generalized here. These were previously derived for orbiting mass m << M. Since Chavda and Chavda primarily deal with m = M , the prior results are presented here in terms of the reduced mass $\mu = mM / (m + M)$ to directly assess limitations on their holeum analysis. The results will be presented in n-space and 3-space. Since they are more general and more precise, they supplant the previous results.

The orbital radius in n-space is

$$r_n = \left[ \frac{j\hbar \pi^{\frac{n-2}{4}}}{[2G_n Mm \mu \Gamma(n/2)]^{1/2}} \right]^{\frac{2}{4-n}}, \qquad (2.1)$$

where j = 1, 2, 3, ... is the principal quantum number. M and m are gravitationally bound masses. The n-space universal gravitational constant $G_n$ changes, in a way that is model dependent, from its 3-space value. The Gamma function $\Gamma(n) \equiv \int_0^\infty t^{n-1} e^{-t} dt$ for all n (integer and non-integer). When n is an integer, $\Gamma(n) = (n-1)!$ $\hbar$ is (Planck's constant)/$2\pi$. For comparison with Chavda and Chavda [1], the masses may be considered black holes with M = m.

In 3-space, equation (2.1) yields

$$r_3 = \frac{j^2 \hbar^2}{GMm\mu} = \frac{2j^2 \hbar^2}{Gm^3} \quad \text{for M = m.} \qquad (2.2)$$

The orbital velocity in n-space is

$$v_n = \left\{ \left[ \frac{2\pi G_n Mm \Gamma\left(\frac{n}{2}\right)}{\pi^{n/2}} \right] \left[ \frac{\mu^{\frac{n-3}{4-n}} \left[ 2G_n Mm \Gamma\left(\frac{n}{2}\right) \right]^{\frac{1}{4-n}}}{(j\hbar)^{2/(4-n)} \pi^{(n-2)/2(4-n)}} \right] \right\}^{1/2}. \qquad (2.3)$$

In 3-dimensions equation (2.3) gives

$$v_3 = \frac{GMm}{j\hbar} = \frac{Gm^2}{j\hbar} \quad \text{for M = m.} \qquad (2.4)$$

Although $v_3$ is independent of $\mu$, $v_n$ is not independent of $\mu$ for higher dimensions. In some holeum cases, m ~ $m_{Planck}$ was used [1], for which equation (2.4) gives $v_3 \approx c$ (the speed of light), necessitating special relativity corrections in their analysis. Smaller masses are also used which reduce $v_3 < c$, but as shown in Sections 4 and 5, this leads to an insufficient binding energy for their chosen realm.

An interesting observation related to the equivalence principle can be made from the acceleration,

$$a_n = \frac{-2\pi G_n M m \Gamma\left(\frac{n}{2}\right)}{\mu \pi^{n/2}} \left[ \frac{[2 G_n M m \mu \Gamma(n/2)]^{1/2}}{j\hbar \pi^{(n-2)/4}} \right]^{\frac{2n-2}{4-n}}. \tag{2.5}$$

In 3-space, equation (2.5) yields

$$a_3 = \frac{-G^3 M^3 m^3 \mu}{(j\hbar)^4} = \frac{-G^3 M^3 m^4}{2(j\hbar)^4} \text{ for } M = m. \tag{2.6}$$

It is interesting to note from equations (2.1) to (2.6) that the acceleration, as well as the orbital radius and velocity, are functions of the mass m in all dimensions as a result of quantization, even when M >> m . Though the presence of m may seem to be an artifact of the Bohr-Sommerfeld condition, the same mass dependency and basically the same results are obtained from the Schroedinger equation. The failure of m to vanish indicates that quantum mechanics is inconsistent with the weak equivalence principle (Rabinowitz, 1990, 2001a). It is not clear whether or not this is related to difficulties encountered in developing a theory of quantum gravity. Classically for M >> m, these variables are independent of the orbiting mass, since m cancels out in accord with the equivalence principle. Furthermore, quantum mechanical interference effects in general, and quantum-gravitational interference effects in particular, depend on the phase which depends on the mass. This is intrinsic to quantum mechanics.

In n-space, the total energy of a gravitationally bound atom is

$$E_n = \frac{(n-4)\mu^{(n-2)/(4-n)}}{n-2} \left[ \frac{G_n M m \Gamma\left(\frac{n}{2}\right)}{\pi^{(n-2)/2}} \right] \left[ \frac{\left[ 2 G_n M m \Gamma\left(\frac{n}{2}\right) \right]^{(n-2)/(4-n)}}{(j\hbar)^{2(n-2)/(4-n)} \pi^{(n-2)^2/2(4-n)}} \right].$$

(2.7)

In 3-space, equation (2.7) reduces to

$$E_3 = -\frac{G^2 M^2 m^2 \mu}{2 j^2 \hbar^2} = -\frac{G^2 m^5}{4 j^2 \hbar^2} \text{ for } M = m. \tag{2.8}$$

Note from equation (2.7) that all energy levels are ≥ 0 in 4 and higher dimensional space, yielding a general result that orbiting bodies in gravitational atoms cannot be bound by energy constraints in higher dimensions, no matter how strong the gravitational attraction, unless short range forces also come into play. (Similarly, for electrostatically bound atoms since they have the same dependence on n.) This may impact string theory if short range forces or other constraints cannot be invoked to achieve stability when the extra dimensions are unfurled.

Mathematically this results from the leading factor $[(n-4)/(n-2)]$ in the complicated quantized equation (2.7). Why $n > 3$ leads to $E_n \geq 0$, can be understood in simpler terms for circular orbits. For a long-range attractive force like gravity with $M \gg m$

$$F_n = \frac{-2\pi G_n M m \Gamma(n/2)}{\pi^{n/2} r_n^{n-1}} = \frac{-mv_n^2}{r_n} \Rightarrow \frac{1}{2} mv_n^2 = \frac{\pi G_n M m \Gamma(n/2)}{\pi^{n/2} r_n^{n-2}}, \qquad (2.9)$$

where the (n-1) exponent of r in $F_n$ results from Gauss' law in n-space, e.g. $F_3 = -GMm/r^2$ because the area of a sphere $\propto r^2$, since we live in a 3-dimensional macroscopic space. Substituting equation (2.9) into the equation for total energy

$$E_n = \frac{1}{2} mv_n^2 + \frac{-2\pi G_n M m \Gamma(n/2)}{(n-2)\pi^{n/2} r_n^{n-2}} = \frac{\pi G_n M m \Gamma(n/2)}{\pi^{n/2} r_n^{n-2}} + \frac{-2\pi G_n M m \Gamma(n/2)}{(n-2)\pi^{n/2} r_n^{n-2}}$$

$$= \left[\frac{n-4}{n-2}\right] \frac{\pi G_n M m \Gamma(n/2)}{\pi^{n/2} r_n^{n-2}} \geq 0 \text{ for } n > 3.$$

(2.10)

This result, with the same prefactor $[(n-4)/(n-2)]$, applies both classically and quantum mechanically since quantization does not change the sign of the co-factor $\frac{\pi G_n M m \Gamma(n/2)}{\pi^{n/2} r_n^{n-2}} > 0$, for positive masses or if both masses are negative.

The same results would be obtained for any other long-range force like the electrostatic force. Short-range forces like the nuclear force are not expected to give this result. It should be noted that Wesson's extension of general relativity to 4-dimensional space (5D space-time) finds a new force[5]. It would be interesting to know if it is possible to have 4-dimensional energetically stable atoms in that 4-space. The results here indicate Euclidean 4-space is singular in that $r_4$ is infinite, and though angular momentum, $L_G = \mu v_4 r_4 = [2\mu G_4 M m / \pi]^{1/2}$ remains finite, $L_G$ and $L_E$ can't be quantized in the usual way because their dependence on $r_4$ vanishes. This and no binding energy for atoms for $\geq$ 4-space has consequences for the 4-space Kaluza-Klein unification of general relativity and electromagnetism, as well as for string theory. In other dimensions, dependence on $r_n$ allows the orbital radius to adjust in quantization of L. Quantization of L, without quantization of r, in 4-space for gravitational and electrostatic atoms leads to interesting results such as possible quantization of $\mu$ etc.

In higher dimensional space the trajectories are generally neither cirlcular, nor elliptical, as the orbits become non-closed curves. Although only circular orbis have been considered, the more complicated central force problem where there is also a radial velocity, yields the same conclusion. Rather than considering $E_n > 0$, we must take into consideration the effective potential energy. The general case can be put in the form of a one-dimensional radial problem in terms of the effective potential energy of the system,

$$V'_n = V_n + L^2 / 2mr^2. \qquad (2.11)$$

$V_n(r)$ is the potential energy, and L is the angular momentum which remains constant because there are no torques in central force motion.

The emitted frequency for closest allowed energy transitions is

$$\nu_n = \frac{\Delta E}{2\pi\hbar}$$

$$= \frac{(n-4)\mu^{(n-2)/(4-n)}}{(n-2)2\pi\hbar} \left[ \frac{G_n Mm \Gamma\left(\frac{n}{2}\right)}{\pi^{(n-2)/2}} \right] \left[ \frac{2G_n Mm \Gamma\left(\frac{n}{2}\right)}{\hbar^{2(n-2)/(4-n)} \pi^{(n-2)^2/2(4-n)}} \right]^{(n-2)/(4-n)} \left[ \frac{1}{(j+2)^{2(n-2)/(4-n)}} - \frac{1}{j^{2(n-2)/(4-n)}} \right] \qquad (2.12)$$

The principal quantum number j changes by an even integer (2, 4, 6, ...) to conserve angular momentum for the system of atom and emitted graviton of spin 2.

In 3-space, equation (2.11) gives the frequency of radiated gravitons

$$\nu_3 = \frac{\Delta E}{2\pi\hbar} = \frac{G_3^2 M^2 m^2 \mu}{4\pi\hbar^3} \left| \frac{1}{j^2} - \frac{1}{(j+2)^2} \right| = \frac{G_3^2 m^5}{8\pi\hbar^3} \left| \frac{1}{j^2} - \frac{1}{(j+2)^2} \right| \qquad (2.13)$$

for M = m.

**3. Viable Black Hole Atoms**

Ordinary gravitational orbits are in the high quantum number, continuum classical limit. In considering GBA, black holes are ideal candidates for the observation of quantization effects [2, 4], since for small orbits very high density matter is necessary. Furthermore, "A little black hole can trap charge internally and/or externally. It could easily trap ~ 10 positive or negative charges externally and form a neutral or charged super-heavy atom-like structure"[3]. Moderately charged black holes could form electrostatically and gravitationally bound atoms. For the present let us consider only gravitational binding where the black hole mass M >> m, the orbiting mass. To avoid complications related to quantum gravity, m can be considered to be made of ordinary matter such as a nucleon or group of bound nucleons.

Newtonian gravity is generally valid for r > 10 $R_H$ since the difference between Einstein's general relativity and Newtonian gravitation gets small in this region. (The black hole horizon or Schwarzschild, radius is $R_H = 2GM/c^2$, where M is the mass of a black hole and c is the speed of light.) This approximation should be classically valid for all scales since the |potential energy|

$$|V| = \frac{G(M)\gamma m}{r} < \frac{G(R_H c^2/2G)\gamma m}{10 R_H} = \frac{\gamma m c^2}{20} \qquad (3.1)$$

is scale independent, where $\gamma = (1 - v^2/c^2)^{-1/2}$. Thus it is necessary that $|V|$ be smaller than 1/20 of the rest energy of the orbiting body of mass m.

We will operate in the realm of Newtonian gravity and thus require the orbital radius $r > 10 R_H$. From equation (2.2) with j =1 and M >> m:

$$r = \frac{\hbar^2}{(GM)m^2} = \frac{\hbar^2}{(R_H c^2/2)m^2} \geq 10 R_H . \qquad (3.2)$$

Solving equation (3.2)

$$R_H \leq \frac{\hbar}{\sqrt{5} mc} = \frac{\lambdabar_C}{\sqrt{5}}, \qquad (3.3)$$

where $\lambdabar_C$ is the reduced Compton wavelength of the orbiting particle. So $r \geq 10 R_H$ is equivalent to the quantum mechanical requirement $\lambdabar_C \geq \sqrt{5} R_H$.

Now let us determine a relationship between M and m that satisfies $r \geq 10 R_H$.

$$r = \frac{\hbar^2}{GMm^2} \geq 10 R_H = 10 \frac{2GM}{c^2} . \qquad (3.4)$$

Equation (3.4) implies that

$$Mm \leq \frac{\hbar c}{\sqrt{20} G} = \frac{(M_{Planck})^2}{\sqrt{20}} . \qquad (3.5)$$

For M = m, r is a factor of 2 larger and equation (34) would yield $M \leq M_{Planck}/\sqrt{10}$. This is why it would be impossible to also avoid the realm of quantum gravity if the two masses are equal as is primarily done in [1].

We can now determine the ground state orbital velocity v in general for any M and m that satisfy $r \geq 10\,R_H$ by substituting equation (3.5) into equation (2.4) for v.

$$v = \frac{G(Mm)}{\hbar} \leq \frac{G}{\hbar}\left(\frac{\hbar c}{\sqrt{20G}}\right) = \frac{c}{\sqrt{20}} = 0.224c. \qquad (3.6)$$

So special relativity corrections would only be small, whereas in some cases $v \approx c$ in [1].

Substituting equation (3.6) for v into equation (2.8), the binding energy is

$$E = -\frac{m}{2}\left[\frac{GMm}{\hbar^2}\right]^2 = -\frac{m}{2}\left[v^2\right] = -\frac{m}{2}\left[\frac{c}{\sqrt{20}}\right]^2 = -\frac{mc^2}{40}. \qquad (3.7)$$

A large range of M >> m can satisfy these equations. For a numerical example, let $m = m_{proton} = 1.67 \times 10^{-27}$ kg. Equation (3.5) implies that $M = 6.36 \times 10^{10}$ kg, with $R_H = 9.43 \times 10^{-17}$m. Equation (3.7) gives a binding energy $E = 3.76 \times 10^{12}$ J = 23.5 MeV, with $v = 6.72 \times 10^7$ m/sec. We want the binding energy E >> kT, so T must be << $2.72 \times 10^{11}$K. Although this is much less than the unification temperature $T_{unif} \sim 10^{29}$ K, and such atoms would not be stable in the very early universe, they could be formed at later times and would be stable over most of the age of the universe. This assumes negligible Hawking radiation [2, 3, 9, 10].

**4. Holeum instability**

Chavda and Chavda [1] propose (p. 2928) that the black holes and the holeum are created, "When the temperature of the big bang universe is much greater than $T_b = mc^2/k_B$, where m is the mass of a black hole and $k_B$ [k here] is the Boltzmann constant...." Let us examine whether the binding energy is great enough to hold holeum together in this high temperature regime. The binding energy between the masses m and m is given by j = 1 in equation (2.8).

In order for the binding energy given by equation (2.8) to be large enough to hold the holeum atom together for high energy collisions in this regime, it is necessary that

$$E_{binding} = |E_{j=1}| = \frac{G^2 m^5}{4\hbar^2} \geq kT >> kT_b = mc^2, \qquad (4.1)$$

where $kT >> kT_b = mc^2$ is given in [1], as quoted above. Equation (4.1) implies that

$$m >> \sqrt{2}\left(\frac{\hbar c}{G}\right)^{1/2} = \sqrt{2}\,m_{Planck}. \qquad (4.2)$$

Equation (4.2) says that masses >> the Planck mass are needed for holeum to be stable in this high temperature regime. This is incompatible with the position in [1, p. 2932] that they are dealing with black holes less than the Planck mass, "In this paper, we consider black holes in the mass range $10^3$ GeV/$c^2$ to $10^{15}$ GeV/$c^2$." This limits the black hole masses from $10^{-24}$ kg to $10^{-12}$ kg, compromising the stability of holeum by tens of orders of magnitude. Both for stability and to circumvent the need for a theory of quantum gravity, masses $\geq 2 \times 10^{-8}$ kg = $m_{Planck}$ are required. But this brings in problems of too small an orbital radius as shown next.

A mass $fm_{Planck} = f(\hbar c/G)^{1/2}$, where f is a pure number can be substituted into equation (2.2) for j = 1 to ascertain the orbital radius, i.e. the separation of the two black holes for the ground state of holeum.

$$r_{j=1} = \frac{2\hbar^2}{G[fm_P]^3} = \frac{2\hbar^2}{G\left[f\left(\frac{2\hbar c}{G}\right)^{1/2}\right]^3} = \frac{2}{f^3 c}\left[\frac{\hbar G}{2c}\right]^{1/2}. \quad (4.3)$$

Let us compare this radius with the black hole $R_H = 2Gm/c^2$ for $m = f(\hbar c/G)^{1/2}$,

$$\frac{r}{R_H} = \frac{2}{f^3 c}\left[\frac{\hbar G}{2c}\right]^{1/2}\left[\frac{c^2}{2Gf(2\hbar c/G)^{1/2}}\right] = \frac{1}{f^4}. \quad (4.4)$$

For a stable orbit, $f = \sqrt{2}$, as determined by equation (4.2). This implies that $r = R_H/4$. This is inconsistent with their use of Newtonian gravity (NG) which requires $r > 2 R_H$ just to avoid collsion between the orbiting black holes. In NG, for equal black hole masses, each LBH orbits at a radius r/2 about the center of mass of the atom. For $r > 10R_H$, NG requires $f < 1/10^{1/4}$ = 0.56, but then the masses are each 0.56 $M_{Planck}$, requiring quantum gravity. For some cases they have $2R_H < 2r < 10R_H$, which is still not adequate.

Higher dimensional atoms will not alleviate this conundrum for the mass or the radius, as shown in Section 2. Section 3 shows that the way out of this problem is to have the mass M be a little black hole which is massive, yet with $R_H \ll r$, and an ordinary orbiting mass $m \ll M_{LBH}$.

For the sake of completeness, we next examine 3-dimensional atoms, and higher dimensional bodies in n-space. The higher dimensional bodies could be nucleons bound by short-range fields like the Yukawa potential. The finding of equation (2.7) that orbiting bodies in atoms cannot be bound by energy constraints for higher than 3-space applies only to long-range fields like the gravitational and electrostatic fields.

**5. Equipartition of kinetic energy in n-space**

*5. 1 Degrees of freedom in n-space*

The total number of degrees of freedom $D_n$ of a d-dimensional body in n-space is

$$D_n = n + (n-1) + (n-2) + \ldots + (n-d), \tag{5.1}$$

for $d \leq n$. Once n coordinates establish the center of mass, there are (n - 1) coordinates left to determine a second reference point on the body, leaving (n - 2) for the third point, ..., and finally (n - d) coordinates for the (d + 1)th reference point.

Since the RHS of equation (5.1) has (d + 1) terms:

$$\begin{aligned} D_n &= n + (n-1) + (n-2) + \ldots + (n-d) = (d+1)\left(\frac{n + (n-d)}{2}\right) \\ &= \left(\frac{d+1}{2}\right)(2n - d). \end{aligned} \tag{5.2}$$

It is interesting to note that $D_n$ is the same for d = (n - 1) and for d = n:

$$D_n(d = n, \text{ or } n-1) = \left(\frac{n+1}{2}\right)(2n - n) = \left(\frac{(n-1)+1}{2}\right)[2n - (n-1)] = \frac{n(n+1)}{2} \tag{5.3}$$

*5.2 Equipartition of kinetic energy*

In 3-space, $D_3$ varies from 3 for d = 0 (point-like object) to 6 for d = 2 (planar object like an ellipse) or d = 3 (object like a spheroid). Choosing n = 10 in reference to string theory, equation (5.2) shows that $D_{10}$ varies from 10 for d = 0 to 55 for d = 9 or 10.

Because the kinetic energy is a quadratic function of velocity in n-space, there will be on the average (1/2)kT of kinetic energy per degree of freedom $D_n$. Let us consider two cases: 1) 3-dimensional body (which could be bound by short range forces) in n-space, i.e. d = 3; 2) n-dimensional body in n-space, i.e. d = n, where n is the number of spatial dimensions in the space-time manifold of (n+1) dimensions.

For a 3-dimensional body in n-space, from equation (5.2) the average kinetic energy is

$$\langle KE \rangle = D_n\left(\tfrac{1}{2}kT\right) = \left(\frac{3+1}{2}\right)(2n - 3)\left(\tfrac{1}{2}kT\right) = (2n - 3)kT \tag{5.4}$$

= 17 kT for n = 10.  [5 kT for a point-like body, depending on scale.]

For an n-dimensional body in n-space, using equation (5.4) gives

$$\langle KE \rangle = D_n (\tfrac{1}{2}kT) = \left(\frac{n+1}{2}\right)(2n-n)(\tfrac{1}{2}kT) = \left(\frac{n(n+1)}{4}\right)kT . \tag{5.5}$$

= 3 kT for n = 3 . [(3/2)kT for a point-like body, depending on scale.]

= (55/2)kT ≈ 28 kT   for n = 10.

Thus in terms of equipartition of energy, at a given temperature T, there can be significantly higher kinetic energy than expected in higher dimensions. Chavda and Chavda [1] are interested in the early universe when the temperature T >> $T_b$ = $mc^2/k$ where m is each black hole mass which makes up holeum.  In 3-space, the average kinetic energy is between 3/2 and 3 $mc^2$ depending on the scale of interaction as to whether the black holes should be considered point-like or 3-dimensional in collisions.  From equations (5.4) and (5.5), in 10-space, holeum (if also bound by short range forces) would be dissociated, since the average kinetic energy is as high as 5 to 28 $mc^2$.

**6. Radiation**

*6.1 Blackbody radiation in n-space*

Let us generalize Boltzmann's derivation of the blackbody radiation law. In n-space, the radiation pressure $P_n = \tfrac{1}{n} u_n$ , where $u_n$ is the energy density.  The internal energy $U_n = u_n V_n$, where $V_n$ is the n-volume.  The thermodynamic relation for internal energy is

$$\frac{\partial}{\partial V_n}(U_n)_T = T\left(\frac{\partial P_n}{\partial T}\right)_V - P_n \Rightarrow \frac{\partial}{\partial V_n}(u_n V_n) = T\frac{\partial}{\partial T}\left(\frac{u_n}{n}\right) - \frac{u_n}{n}. \tag{6.1}$$

Equation (6.1) leads to

$$\frac{du_n}{u_n} = (n+1)\frac{dT}{T} \Rightarrow u_n \propto T^{n+1}. \tag{6.2}$$

Thus the n-dimensional equivalent of the Stefan-Boltzmann blackbody radiation law from equation (6.2) is

$$P_{BBn} \propto cu_n \propto T^{n+1} , \tag{6.3}$$

It is interesting to note that the dimensionality of macroscopic space can be determined by measuring the exponent of the blackbody radiation law. If energetically stable atoms (e.g. bound by additional short-range forces) could exist in (n > 3)-space, equation (6.3) says that for a given T, the collective blackbody radiation of these atoms emits considerably higher power than in 3-space.

*6.2 Hawking radiation in n-space*

The Hawking radiation power, $P_{SH}$, follows from the Stefan-Boltzmann blackbody radiation power/area law $\sigma T^4$ for black holes. For Hawking [6, 7] :

$$P_{SH} \approx 4\pi R_H^2 \left[\sigma T^4\right] = 4\pi \left(\frac{2GM}{c^2}\right)^2 \sigma \left[\frac{\hbar c^3}{4\pi k GM}\right]^4 = \frac{\hbar^4 c^8}{16\pi^3 k^4 G^2} \{\sigma\} \left[\frac{1}{M^2}\right] \quad (6.4)$$

where $\sigma$ is the Stefan-Boltzmann constant. To avoid the realm of quantum gravity, Hawking requires the black hole mass M > $M_{Planck}$.

Since Hawking radiation [6, 7] was derived as blackbody radiation from a black hole, using equation (6.3), $R_{Hn}$ and $T_n$ from [2], the Hawking power radiated in n-space for n ≥ 3:

$$P_{SHn} \propto \left[R_{Hn}\right]^{n-1} \left[T_n\right]^{n+1} \propto \left[M^{1/(n-2)}\right]^{n-1} \left[M^{-1/(n-2)}\right]^{n+1} \propto \frac{1}{M^{2/(n-2)}}$$
$$\propto M^{-2} \text{ for } 3-\text{space}. \quad (6.5)$$
$$\propto M^{-1/4} \text{ for } 10-\text{space}.$$

Although ordinary blackbody radiation is dramatically large $\propto T^{11}$ in 10-space, the mass dependency of Hawking radiation decreases for dimensions higher than 3 for LBH.

*6.3 Compact dimensions attenuate Hawking radiation*

Another approach assumes the correctness of the Hawking model, but analyzes the effects of additional compact dimensions on the attenuation of this radiation. Argyres et al [8] conclude that the properties of LBH are greatly altered and LBH radiation is considerably attenuated from that of Hawking's prediction. Their LBH are trapped by branes so essentially only gravitons can get through the brane (which may be thought of as an abbreviation for vibrating membrane). For them, not only is the radiation rate as much as a factor of $10^{38}$ lower, but it also differs by being almost entirely gravitons.

*6.4 Gravitational tunneling radiation (GTR)*

Gravitational tunneling radiation (GTR) may be emitted from black holes in a process differing from that of Hawking radiation, $P_{SH}$, which has been undetected for three decades. Belinski [9], a noted authority in the field of general relativity, unequivocally concludes "the effect [Hawking radiation] does not exist." GTR is offered as an alternative to $P_{SH}$. In the gravitational tunneling model [3], beamed exhaust radiation tunnels out from a LBH with radiated power, $P_R$, due to the field of a second body, which lowers the LBH gravitational potential energy barrier and gives the barrier a finite width. Particles can escape by tunneling (as in field emission). This is similar to electric field emission of electrons from a metal by the application of an external field.

Although $P_R$ is of a different physical origin than Hawking radiation, we shall see that it is analytically of the same form, since $P_R \propto \Gamma P_{SH}$, where $\Gamma$ is the transmission probability approximately equal to the WKBJ tunneling probability $e^{-2\Delta\gamma}$ for LBH. The tunneling power [4] radiated from a LBH for $r \gg R_H$ is:

$$P_R \approx \left| \frac{\hbar c^6 \langle e^{-2\Delta\gamma} \rangle}{16\pi G^2} \right| \frac{1}{M^2} \sim \frac{\langle e^{-2\Delta\gamma} \rangle}{M^2} \left[ 3.42 \times 10^{35} W \right], \qquad (6.6)$$

where M in kg is the mass of the LBH. No correction for gravitational red shift needs to be made since the particles tunnel through the barrier without change in energy. The tunneling probability $e^{-2\Delta\gamma}$ is usually $\ll 1$ and depends on parameters such as the width of the barrier, M, and the mass of the second body [3].

Hawking invoked blackbody radiation in the derivation of equation (6.4). But it was not invoked in the GTR derivation of equation (6.6).[3] Although $P_R$ and $P_{SH}$ represent different physical processes and appear quite disparate, the differences in the equations almost disappear if we substitute into equation (6.4) the value obtained for the Stefan-Boltzmann constant $\sigma$ by integrating the Planck distribution over all frequencies:

$$\sigma = \left\{ \frac{\pi^2 k^4}{60 \hbar^3 c^2} \right\}, \qquad (6.7)$$

$$P_{SH} = \frac{\hbar^4 c^8}{16\pi^3 k^4 G^2} \left\{ \frac{\pi^2 k^4}{60 \hbar^3 c^2} \right\} \left[ \frac{1}{M^2} \right] = \frac{\hbar c^6}{16\pi G^2} \left\{ \frac{1}{60} \right\} \left[ \frac{1}{M^2} \right]. \qquad (6.8)$$

Thus $P_R = 60 \langle e^{-2\Delta\gamma} \rangle P_{SH}$. \qquad (6.9)

GTR is beamed between a black hole and a second body, and is attenuated by the tunneling probability $\langle e^{-2\Delta\gamma} \rangle$ compared to $P_{SH}$. Leaving aside the attenuation factor, it is not clear if there is physical significance to the same analytic form for $P_R$ and $P_{SH}$. It could simply result from the dimensionality requirement that they are both in units of power.

Two LBH may get quite close for maximum GTR. In this limit, there is a similarity between GTR and what is expected from the Hawking model. GTR produces a repulsive recoil force between two bodies due to the beamed emission between them. Since the tidal forces of two LBH add together to give more radiation at their interface in his model, this also produces a repulsive force.

*6.5 Diminished Hawking radiation from charged black holes*

As a black hole becomes more and more charged, the Hawking radiation decreases until in the limit of maximum charge containment there is none. In 1986, Balbinot [10] demonstrated that highly charged black holes do not Hawking radiate. He determined that "For an extreme Reissner-Nordstrom [highly charged] black hole ... there is no Hawking evaporation." There is no mention of this in [1], nor do they consider moderately charged black holes which could form electrostatically and gravitationally bound atoms. Though charged black holes have not been candidates for dark matter, charged black hole atoms, neutralized by orbiting charges, have been considered [3].

*6.6 Gravitational radiation*

An advanced quadrupole suspension design [11] for the U.S. Laser Interferometer Gravitational-Wave Observatory (LIGO) has recently been described to measure gravitational radiation from distant sources. Preparation is also being made by other teams around the world. In addition to LIGO, there is VIRGO (France/Italy); GEO-600 (Britain/ Germany); TAMA (Japan); and ACIAGA (Australia). The detectors are laser interferometers with a beam splitter and mirrors suspended on wires. The predicted gravitational wave displaces the mirrors and shifts the relative optical phase in two perpendicular paths. This causes a shift in the interference pattern at the beam splitter. It is expected that by 2010, the devices will be sensitive enough to detect gravitational waves up to $10^2$ Megaparsecs ( $3.26 \times 10^8$ lightyear = $3.1 \times 10^{24}$ m). A challenge arose because the detector noise does not satisfy the usual assumptions that it be stationary and Gaussian [12].

As shown in Section 2, quantized gravitational radiation is possible from GBA. The possibility was presented that a signal from such potentially nearby sources can compete or interfere with distant sources such as neutron stars, binary pulsars, and coalescing black holes [2]. Signals from such sources are expected to have frequencies in the range from 10 Hz to $10^4$ Hz [13]. It was shown that gravitational radiation from orbital de-excitation of an ordinary mass orbiting a LBH would have a detectable frequency ~ $10^3$ Hz [2]. A mass m ~ $10^-$ $10^{27}$ kg orbiting a LBH of mass M ~ 10 kg, would emit a frequency ~ $10^3$ Hz. in going from the j =3 state to the j = 1 ground state.

**7. Different views of black holes as dark matter candidates**

Discovery of the nature of dark matter will help to define what the universe is made of. It will reveal the invisible particles carrying the gravitational glue that holds the universe, galaxies, and clusters of galaxies together, and determines the curvature of space. We should not arbitrarily rule out the possibility that dark matter can occasionally manifest itself on earth. In addition to the Chavda and Chavda view [1] considered throughout this paper, other views of black holes as dark matter candidates are now presented to give a broader perspective.

*7.1 Large black holes: $10^{14}$ kg ≤ $M_{BH}$ ≤ $10^{36}$ kg*

A review article of 1984 [14 and references therein] presents the prevailing view of black holes as constituents of dark matter. The article considers only rather massive black holes as a possible component of dark matter: "A third cold DM [dark matter] candidate is black holes of mass $10^{-16}$ $M_{sun}$ ≤ $M_{BH}$ ≤ $10^6$ $M_{sun}$, the lower limit implied by the non-observation of γ rays from black hole decay by Hawking radiation...." ($M_{sun}$ = 2 x $10^{30}$ kg.)

*7.2 Medium black holes: $10^{12}$ kg ≤ $M_{BH}$ ≤ $10^{30}$ kg*

Trofimenko in 1990 [15] discussed the possibility that black holes up to the mass of the sun, $M_{sun}$, are involved in a multitude of geophysical and astrophysical phenomena such as in stars, pulsars, and planets. Although he did not explicitly consider them as candidates for dark matter, for him they are "universal centres of all cosmic objects." That makes them such candidates implicitly. He was not concerned with the ramifications of LBH radiation, nor the time for LBH to devour their hosts. His lower mass limit of $10^{12}$ kg comes from the failure to detect Hawking radiation, and expected smallest primordial mass survival.

*7.3 Primordial black holes: $M_{BH}$ ~ $10^{13}$ kg*

Beginning in 1993, Alfonso-Faus [16] proposed "primordial black holes, massive particles about $10^{40}$ times the proton mass" [$10^{40}(10^{-27}$ kg) = $10^{13}$ kg] as his dark matter candidate. He goes on to say that they do not radiate by Hawking radiation, but does not comment on how they radiate. Elsewhere [17] he asserts a radiation wavelength of $10^8$ cm from black holes that is the geometric mean

-15-

between the radius of such a primordial black hole ($10^{-12}$ cm) and the radius of the universe ($10^{28}$ cm). With such a long wavelength, he concludes that they radiate, "about $10^{40}$ times lower " than in the Hawking model and hence "they would still be around....."

*7.4. Higher Dimensional Primordial Black Holes: $10^{29}$ kg ≤ M ≤ $10^{34}$ kg*

Argyres et al [8] examine primordial black holes (PBH) in higher compact dimensions. They conclude that for 6 extra compact dimensions (9-space), 0.1 solar mass PBH are dark matter candidates, but that this increases to ~$10^4$ solar masses if there are only 2 or 3 extra dimensions (5 to 6-space). Smaller PBH might be expected, since for them PBH radiation is almost entirely gravitons. In standard Hawking radiation from LBH, > MeV photons would dissociate big bang nucleosynthesis products, devastating the presently propitious predictions of light element abundances. They conclude, "The lightest black holes that can be present with any significant number density in our universe today are thus formed immediately after the epoch of inflationary reheating."

*7.5 Primordial little black holes: $10^{-7}$ kg ≤ $M_{LBH}$ ≤ $10^{19}$ kg*

Starting in 1998, it was proposed that black holes radiate by GTR allowing primordial LBH to be regarded as candidates for the dark matter of the universe [2,3, 18]. These were the smallest masses ($10^{-7}$ kg to $10^{19}$ kg) considered until 2002. Since GTR can be greatly attenuated compared with Hawking radiation, cf. Section 6.4, this has strong implications down to the smallest masses of LBH, whether the LBH are free or GBA. For Hawking [7], the smallest LBH that can survive to the present is M ~ $10^{12}$ kg . Let us see what GTR predicts.

The evaporation rate for a black hole of mass M is $d(Mc^2)/dt = -P_R$, which gives the lifetime

$$t = \frac{16\pi G^2}{3\hbar c^4 \langle e^{-2\Delta\gamma} \rangle} \left[M^3\right]. \quad (7.1)$$

This implies that the smallest mass that can survive up to a time t is

$$M_{small} = \left(\frac{3\hbar c^4 \langle e^{-2\Delta\gamma} \rangle}{16\pi G^2}\right)^{1/3} \left[t^{1/3}\right]. \quad (7.2)$$

Primordial black holes with M >> $M_{small}$ have not lost an appreciable fraction of their mass up to the present. Those with M << $M_{small}$ would have evaporated away long ago.

Thus the smallest mass that can survive within $\sim 10^{17}$ sec (age of our universe) is

$$M_{small} \geq 10^{12} \left\langle e^{-2\Delta\gamma} \right\rangle^{1/3} \text{ kg}. \tag{7.3}$$

Hawking's result [6, 7] of $10^{12}$ kg is obtained by setting $e^{-2\Delta\gamma} = 1$ in eq. (7.3). Since $0 \leq e^{-2\Delta\gamma} \leq 1$, an entire range of black hole masses much smaller than $10^{12}$ kg may have survived from the beginning of the universe to the present than permitted by Hawking's theory.

For example, if the average tunneling probability $\left\langle e^{-2\Delta\gamma} \right\rangle \sim 10^{-45}$, then $M_{small} \sim 10^{-3}$ kg. For $M_{univ} \sim 10^{53}$ kg, $V_{univ} \sim 10^{79}$ m$^3$ (radius of $15 \times 10^9$ light-year = $1.4 \times 10^{26}$ m), the average density of such LBH would be 1 LBH per $10^{23}$ m$^3$. The velocity of our local group of galaxies with respect to the microwave background (cosmic rest frame), $v_{LBH} \sim 6.2 \times 10^5$ m/sec [19], is a reasonable velocity for LBH with respect to the earth. This may make it possible to detect their incident flux $\sim (10^{-23}/\text{m}^3)(6.2 \times 10^5 \text{ m/sec}) \sim 10^{-17}/\text{m}^2\text{sec}$ on the earth. [2]

*7.6 Non-radiating holeum : $10^{-24}$ kg $\leq M_{BH} \leq 10^{-12}$ kg*

In 2002 Chavda and Chavda [1] introduced a novel proposal that gravitationally bound black holes will not Hawking radiate by analogy to the neutron. Their model has been analyzed in this paper. It appears from this analysis that stable holeum cannot exist in 3-space, or in any higher dimensions. Therefore whether or not such an object might Hawking radiate is a moot point.

The analogy between holeum and a bound neutron may not apply. A neutron in free space decays with a half-life of about 10.6 minutes. The neutron spontaneously decays into a proton, an electron, and an antineutrino. This is energetically possible because the neutron's rest mass is greater than that of the decay products. This difference in rest mass manifests itself in an energy release of $1.25 \times 10^{-13}$ J (0.782 MeV). The situation in a nucleus is complicated by many factors such as Fermi levels of the neutrons and the protons, etc. Neutrons do decay in nuclei that are beta emitters despite their relatively large binding energy which is typically 1 to $1.4 \times 10^{-12}$ J (6 to 8 MeV). Other than the interesting neutron analogy, they give no compelling reasons for the absence of Hawking radiation in black hole GBA.

Most of the orbital radii are in the strong field region $2 R_H < r < 10 R_H$, requiring general relativity corrections. Therefore in the absence of $r > 10 R_H$, their use of Newtonian gravity is questionable. There is an error by a factor of $10^2$ too high in the orbital radius given by their equation (45).

In considering little black hole masses $M_{LBH} < M_{Planck} \sim 10^{-8}$ kg, their analysis [1] exceeds another domain of validity which requires a theory of quantum gravity. For larger masses and larger radii than they use, it would be easy to agree with their choice of quantized Newtonian Gravity. Except for the 0 angular momentum state (which does not exist semi-classically), essentially the same results are obtained semi-classically and quantum mechanically. It is generally agreed that for $M_{LBH} > M_{Planck}$ one may describe LBH with semi-classical physics, and quantum gravity is needed for $M_{BH} \leq 10^{-8}$ kg, since this is below the Planck scale where a little black hole has $R_H \leq 10^{-35}$ m.

Neglect of special relativity is a further problem in [1], since in some cases the orbital velocity in holeum $v \approx c$. It is relevant to note that non-relativistic quantum mechanics and even the semi-classical Bohr-Sommerfeld equation give accurate energy levels for hydrogen despite being non-relativistic. This is because they neglect the serendipitously near-canceling effects of both relativity and spin. One is the relativistic increase of the electron's mass as its velocity increases near the proton. The other is the interaction of the electron's intrinsic magnetic moment with the Coulomb field of the proton. Since a neutral LBH has no magnetic moment, there are no canceling effects and one may expect a much less reliable result from a treatment that neglects special relativity.

Although some of these problems were recognized and discussed, no attempt to ameliorate them was made[1]. Their primary choice of equal masses was likely motivated by the deuteron where the neutron becomes stable in binding with a proton. However as shown in Section 4, such a choice forces these masses to be less than the Planck mass. As shown in Section 3, all the above problems can be eliminated by having a black hole of mass $M > M_{Planck}$ and an ordinary mass $m \ll M$, with supressed Hawking radiation (cf.Sec.6).

**8. New black hole entropy**

Since it is generally accepted that black hole radiation is independent of the time-history of the black hole formation, we may for theoretical purposes consider a black hole that has existed for an infinite length of time. Conclusions about the entropy of such a static everlasting black hole should also be valid for all black holes, despite the virtual inconsistency that any black hole would have evaporated away after an infinite time. This should be legitimate for *gedanken* purposes. The orthodox view is that such a black hole should be empty inside the horizon, except for a singularity at its center.

To possibly gain an insight into the entropy of black holes, let us model the black hole of mass M as a spherical gravitational bottle filled with an ideal gas of N identical particles, each of mass m. If the results look promising, the complexity of this model of a single species in a gravitational equipotential inside the hole can be further increased. We start with the ideal gas law in terms of the mass density $\rho$ of the black hole

$$P = \left(\frac{N}{V}\right)kT = \left(\frac{\rho}{m}\right)kT, \qquad (8.1)$$

where P is the pressure, $V = \left(\frac{4\pi}{3}\right)R_H^3$ is the volume, and T is the temperature. The density of the black hole is

$$\rho = \frac{M}{\left(\frac{4\pi}{3}\right)R_H^3} = \frac{M}{\left(\frac{4\pi}{3}\right)\left(\frac{2GM}{c^2}\right)^3} = \frac{3c^6}{32\pi G^3 M^2}. \qquad (8.2)$$

For the black hole temperature Hawking's 1975 value [7]:

$$T = \left[\frac{\hbar c^3}{4\pi kG}\right]\frac{1}{M}. \qquad (8.3)$$

Combining eqs. (8.2) and (8.3) we have

$$T = \left[\frac{\hbar c^3}{4\pi kG}\right]\frac{1}{M} = \frac{\hbar}{k}\sqrt{\frac{2\rho G}{3\pi}}. \qquad (8.4)$$

Combining eqs. (8.1) and (8.4), we obtain the pressure inside the black hole modeled as a gravitational bottle

$$P = \left(\frac{\rho}{m}\right)k\left[\frac{\hbar}{k}\sqrt{\frac{2\rho G}{3\pi}}\right] = \frac{\hbar}{m}\left(\frac{2G}{3\pi}\right)\rho^{3/2}, \qquad (8.5)$$

containing $N = M/m$ gravitationally bound particles, each of mass m. From kinetic theory, the pressure is also

$$P \approx \tfrac{1}{3}(N/V)\left[\overline{mv^2}\right] \approx \tfrac{1}{3}\rho c^2. \qquad (8.6)$$

Equating eq. (8.5) with eq. (8.6) and solving for the density $\rho$

$$\rho = \left(\frac{3\pi}{2G}\right)\left(\frac{m^2 c^4}{9\hbar^2}\right) = \left(\frac{N}{V}\right)m. \qquad (8.7)$$

Combining eqs.(8.2) and (8.7) we find the number of particles inside the black hole is

$$N = \left(\frac{3}{4\pi}\right)\left(\frac{m_P}{m}\right)^2 = \left(\frac{4\pi}{3}\right)\left(\frac{M}{m_P}\right)^2, \qquad (8.8)$$

where the Planck mass $m_P = (\hbar c/G)^{1/2}$.

For large N, the entropy of the black hole is

$$S_{bh} = k\ell n(e^N) = kN = k\left(\frac{4\pi}{3}\right)\left(\frac{M}{m_P}\right)^2 = \frac{kAc^3}{12\hbar G}. \qquad (8.9)$$

This is 1/3 as large as the Bekenstein black hole entropy

$$S_{Bek} = \frac{kAc^3}{4\hbar G} = 3S_{bh}. \qquad (8.10)$$

The area of a black hole in terms of its density is

$$A = 4\pi R_H^2 = \frac{3M}{\rho R_H} = \frac{3M}{\rho\left(\frac{2GM}{c^2}\right)} = \frac{3c^2}{2\rho G}. \qquad (8.11)$$

Hence in terms of mass density, the entropy of a black hole is

$$S_{bh} = kAc^3 / 4G\hbar = \frac{kc^3}{4G\hbar} 4\pi R_H^2 = \frac{kc^3}{4G\hbar}\left(\frac{3c^2}{2\rho G}\right) = \frac{3kc^5}{8\rho G^2 \hbar}. \qquad (8.12)$$

The pressure equation (8.5) may also be written as

$$P = \left(\frac{\rho}{m}\right) k \left[\frac{\hbar}{k}\sqrt{\frac{2\rho G}{3\pi}}\right] = \frac{\hbar}{m}\left(\frac{2G}{3\pi}\right)\left[\frac{3c^6}{32\pi G^3 M^2}\right], \qquad (8.13)$$

indicating that the black hole pressure gets very high as its mass gets low.

## 9. Conclusion

Examination of other radiation attenuation alternatives and dark matter models indicates a number of ways in which Hawking radiation may not be a deterrent to little black holes as dark matter candidates. Blackbody and Hawking radiation were analyzed in n-space, with the noteworthy result that even though blackbody radiation is dramatically larger in higher dimensions, Hawking radiation decreases for dimensions higher than 3 for little black holes.

    An incompatibilty between quantum mechanics and the weak equivalence principle was demonstrated in Section 2. Perhaps this can shed light on why attempts to develop a theory of quantum gravity have led to discrepancis and even contradictions.

    Orbits in n-space were analyzed to see if higher dimensions could enhance the stability of gravitationally bound atoms. Instead, it was found that orbiting bodies in higher dimensional gravitational atoms cannot be bound by energy constraints in higher dimensions (elliptical orbits would not change this conclusion). Similarly for electrostatically bound atoms. This is because there is no binding energy for n > 3, no matter how strong the coupling between the two bodies. So even in the early universe when the gravitational force is very high because all the fundamental forces have the same strength, in higher dimensions

there would be no gravitational or electrostatic atoms unless short-range forces come into play. This has ramifications for both Kaluza-Klein theory and string theory.

This paper questions the domains of validity in [1]. Aside from this issue, it was also shown that even with the high binding energy of holeum, it is not enough to remain stable in collisions for their stated condition that $kT \gg kT_b = mc^2$ in 3-space [1]. This was done in general by deriving the degrees of freedom of a d-dimensional body in n-dimensions (cf. Section5), and applying the equipartition of kinetic energy. Nevertheless, Chavda and Chavda are to be applauded for the novelty of their analogy between holeum and a bound neutron to determine if this can solve a most difficult and important problem in astrophysics.

A promising new approach to black hole entropy was presented that results in a value of black hole entropy 1/3 the magnitude of Bekenstein's value.

**Acknowledgment**

I would like to thank Mark Davidson and Steve Crow for helpful discussions, and Professor Albert W. Overhauser for bringing the Chavda and Chavda paper to my attention.